\documentclass[twoside]{article}
\usepackage{fleqn,espcrc2,epsfig}

\usepackage{graphicx}
\usepackage[figuresright]{rotating}

\newcommand{\AmS}{{\protect\the\textfont2
  A\kern-.1667em\lower.5ex\hbox{M}\kern-.125emS}}

\title{Future Atmospheric Neutrino Detectors}

\author{A. Geiser\address[UniHH]{II. Inst. f. Experimentalphysik, 
        Hamburg University, Luruper Chaussee 149, 
        D-22761 Hamburg, Germany. \quad e-mail: Achim.Geiser@desy.de}}
       
\begin{document}

\begin{abstract}
Future experiments focusing on atmospheric neutrino detection are 
reviewed. One of the main goals of these experiments is the detection of an 
unambiguous oscillation pattern ($\nu_\mu$ reappearance) to prove the 
oscillation hypothesis. 
Further goals include the discrimination of 
$\nu_\mu - \nu_\tau$ and $\nu_\mu-\nu_{\rm sterile}$ oscillations, and
the detection of a potential small $\nu_\mu - \nu_e$ contribution.
The search for matter effects in  three or more flavour oscillations can be 
used to constrain hybrid oscillation models and potentially
measure the sign of $\delta m^2$.
The detectors and measurement techniques proposed to achieve these goals are 
described, and their physics reach is discussed.

\vspace{1pc}
\end{abstract}

\maketitle

\section{Introduction}

At the Neutrino '94 conference in Eilat, Israel, the title of the summary talk
on atmospheric neutrinos \cite{Barish} was: ``Is There an Atmospheric 
Neutrino Anomaly?''. Already then, B.C. Barish concluded that ``it seems 
strongly indicative that the anomaly 
is indeed real, and that it is due to a deficit in the number of observed 
muon neutrinos.''    
This conclusion was beautifully confirmed by the presentation of 
``Evidence for $\nu_\mu$ oscillations'' by Super-Kamiokande \cite{SK-98}  
in Takayama, Japan, 4 years later \cite{SK-Takayama}, implying the disappearance 
of muon neutrinos into some other neutrino flavour.

In the two flavour approximation, the survival probability for
muon neutrinos in vacuum can be expressed by the well known
oscillation formula
\begin{equation}
\label{prob}
P(L/E) = 1-\sin^2(2\Theta)\ \sin^2(1.27\ \Delta m^2\ L/E)
\end{equation}
where $L$ is the distance travelled in km, $E$ is the neutrino energy
in GeV, $\Theta$ is the neutrino mixing angle, and $\Delta m^2$ is the
difference of the mass square eigenvalues expressed in eV$^2$.
This simple relation could be modified by a contribution of additional
flavours, and by matter effects \cite{matter} which depend on the     
neutrino flavour.

Oscillations into $\nu_\tau$ are the currently preferred hypothesis.
A large contribution of oscillations into $\nu_e$ is excluded by the CHOOZ
reactor results \cite{CHOOZ}, and recent indications from Super-Kamiokande 
strongly disfavour pure $\nu_\mu - \nu_{\rm sterile}$ oscillations
\cite{SK-nu2000}.
Hybrid oscillations into both $\nu_\tau$ and $\nu_{\rm sterile}$, and a small 
contribution of $\nu_\mu - \nu_e$ oscillations are however  still allowed
\cite{Lisi}.  
Furthermore, due to the limited 
L/E resolution, exotic alternatives to the oscillation hypothesis such as 
neutrino decay \cite{decay}, large extra dimensions 
\cite{barbieri} or neutrino decoherence \cite{decoherence} remain fully
valid explanations \cite{Lisi}.

Over the last few years, an impressive effort has been made towards the
further clarification of the atmospheric neutrino anomaly with the help
of long baseline neutrino beams in Japan \cite{K2K,K2K-nu2000}, 
the US \cite{MINOS,MINOS-nu2000}
and Europe \cite{CNGS,OPERA,ICANOE,LNGS-nu2000}.      
This program is starting to yield first
results \cite{K2K-nu2000}.
Here, we would like to present the case why these measurements need to be 
complemented by further, more precise atmospheric neutrino measurements,
and how these measurements can be achieved. 

\subsection{Why atmospheric neutrinos?}

Atmospheric neutrino experiments offer several advantages over
currently operational or planned long baseline neutrino beam programs.
\begin{itemize}
\item A very large $L/E$ range (from about 1 km/GeV to 10$^5$ km/GeV;
a typical long baseline beam covers only one or two orders of
magnitude).  Therefore, a very large range of oscillation parameters
can be studied simultaneously.
\item Two identical sources for a single detector:
      a near (downgoing neutrinos) and a far (upgoing neutrinos) one.
\item For some of the measurements, e.g. the confirmation of the oscillation
      pattern, there is currently no alternative to atmospheric neutrino 
      detectors if the atmospheric $\Delta m^2$ is low. The pattern 
      measurement is sensitive to test and exclude alternative 
      explanations and is competitive to long baseline studies even at high 
      $\Delta m^2$.
\item During the next decade large matter effects
with high energy neutrinos can only be observed in atmospheric
neutrino experiments, since the current long baseline distances 
are too short for a significant effect.  
The measurement of the presence or absence of such effects can significantly
constrain complicated hybrid oscillation models. The measurement of 
a difference in the neutrino and antineutrino oscillation parameters might 
allow the measurement of the sign of $\Delta m^2$, i.e. the neutrino mass 
hierarchy (matter effects are expected to dominate over CP violation 
effects).
\end{itemize}

\section{Future atmospheric neutrino detectors}

Most future detectors able to make significant contributions to the detection
of atmospheric neutrinos are multi-purpose detectors which also cover many
other physics topics, like the detection of neutrinos from long baseline beams,
detection of solar neutrinos, supernova neutrinos, and nucleon decay,
or the study of cosmic ray muons and other astrophysics topics.

The reference against which future improvements should be evaluated is 
given by the results of Super-Kamiokande \cite{SK-98,SK-nu2000},  a 50 kt 
(23 kt fiducial) water Cherenkov detector.

\subsection{Water Cherenkov detectors: UNO, AQUA-RICH} 

Some of the atmospheric neutrino measurements (e.g. $\nu_\tau$/$\nu_s$ 
separation) are still statistically limited, and could be improved by 
extending the Super-Kamiokande concept to a larger detector mass.

\begin{figure}[ht]
\epsfig{file=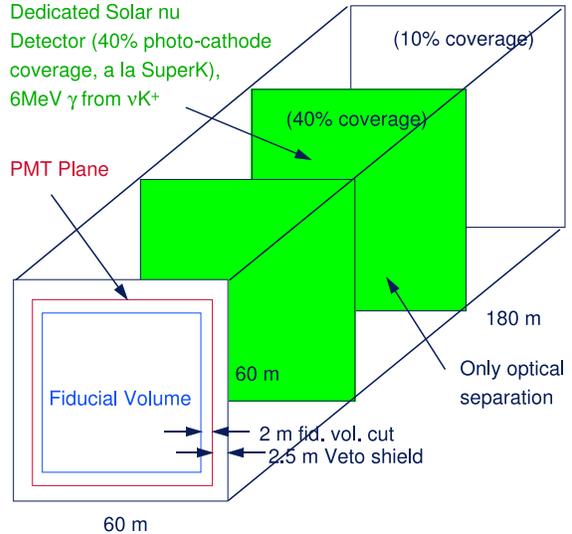,width=8cm}
\vspace{-1.5cm}
\caption{The UNO detector concept.}
\vspace{-.5cm}
\label{fig:UNO}
\end{figure}

The concept of a 650 kt (450 kt fiducial) water Cherenkov detector, 
tentatively called UNO \cite{UNO}
(Ultra underground Nucleon decay and neutrino Observatory detector),
is currently being discussed (fig. \ref{fig:UNO}). 
It would be subdivided into three cubic compartments of 
$60 \times 60 \times 60$ m$^3$. The walls of the central compartment would 
be equipped with photomultipliers ``a la Super-K'', to be fully sensistive 
also to low energy (MeV) neutrinos. A less dense photomultiplier coverage
is forseen for the two outer compartments, dedicated to higher energy events.  
Apart from Super-K-like measurements with 20-fold increased statistics, 
the $\tau$ production rate from $\nu_\mu - \nu_\tau$ oscillations might be
large enough in such a detector to extract an explicit $\tau$ appearance 
signal from atmospheric neutrinos \cite{UNO}. Optionally, an external muon
momentum and charge identifier could be addeed.  

\begin{figure}[htb]
\epsfig{file=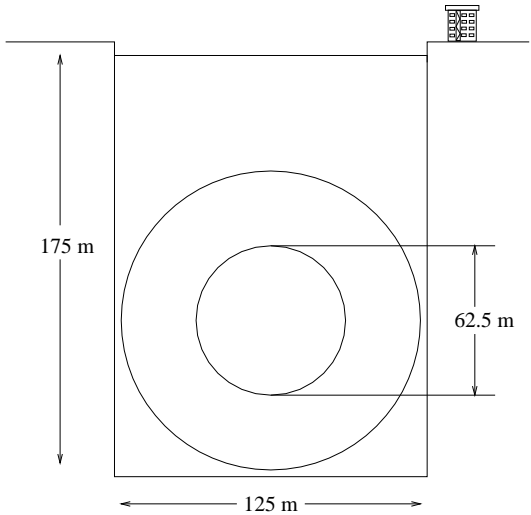,width=7cm}
\vspace{-.8cm}
\caption{The AQUA-RICH detector concept.}
\vspace{-0.5cm}
\label{fig:aqua}
\end{figure}

A different detector concept, based on the Ring Imaging Cherenkov (RICH) 
technique, is envisaged by the AQUA-RICH project \cite{AQUA-RICH}.
This project is currently in the R\&D phase (fig. \ref{fig:aqua}).      
Two spherical detector structures equipped with Hybrid Photo-Detectors
(HPD's) are immersed into a large water tank (fig. \ref{fig:aqua}).      
The upper part of this tank is used as a shield against low energy cosmic 
rays. The inner surface of the outer sphere is equipped as a mirror to focus
the Cherenkov light onto the outer surface of the inner sphere. 
This inner sphere (dome) is densely covered with HPD's to detect the focused 
Cherenkov rings. In addition, the mirror sphere is sparsely equipped with
additional HPD's to also directly detect the non-focused Cherenkov light. 
A fiducial mass of 1 Mt is envisaged for this project.

While the mirror HPD's can be used to reconstruct the neutrino interaction
vertex and the particle flight direction, the space-time structure of the 
Ring Image on the dome HPD's yields a measurement of the particle momentum 
through multiple scattering. A momentum resolution of 7\% can be achieved
\cite{Kai}.
This yields an L/E resolution which is good enough to resolve the sinusoidal
oscillation pattern \cite{AQUA-RICH} over the complete Super-Kamiokande
allowed range.

\subsection{Magnetised tracking calorimeters: MONOLITH}

The water detectors discussed above have the advantage of a very large mass,
and therefore large statistics, but two main drawbacks. The time scale
for their realization spans at least 10 years, and the implementation of a
magnetic field (if any) to measure the muon charge is complicated and costly.
A different detector concept is being followed using large magnetised tracking 
calorimeters. 

The advantage of such detectors is that they can easily measure the muon 
charge, and therefore separate the neutrino and antineutrino components.
In the context of atmospheric neutrinos, this is very useful to study 
matter effects, which can be significantly different in the two cases.
Furthermore, the energy of the hadronic system and the momentum of 
semi-contained muons can be measured. This yields a good reconstruction
of the neutrino energy up to the highest energies. At these energies 
the neutrino angular resolution, which determines the resolution on L,
is also good. Hence, very good L/E resolution can be obtained
\cite{goodLE},
which is needed to resolve the oscillation pattern.

Two such detectors, the approved MINOS detector at Fermilab \cite{MINOS} 
and the NOE part of the proposed ICANOE detector at Gran Sasso \cite{ICANOE}, 
have been primarily designed as long baseline beam detectors. They are 
described in other contributions to this conference 
\cite{MINOS-nu2000,LNGS-nu2000}. Their small fiducial mass (3.5 kt or less)
severely limits the statistics, and therefore their potential contribution
to atmospheric neutrino measurements.

\begin{figure}[t]
\epsfig{file=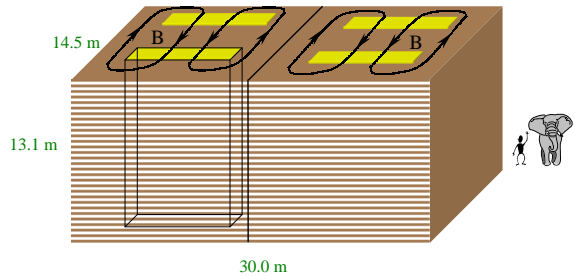, angle=-90, width=7.5cm}
\vspace{-1cm}
\caption{Schematic view of the MONOLITH detector \cite{MONOLITH}.
The arrangement of the magnetic
field is  also shown.} \label{fig:modetect}
\vspace{-.5cm}
\end{figure}

MONOLITH (Massive Observatory for Neutrino Oscillations or LImits on 
THeir existence),
a 34 kt magnetised iron detector dedicated to the measurement of
atmospheric neutrinos in the Gran Sasso lab in Italy, has been officially 
proposed this summer \cite{MONOLITH}.  Its fiducial mass of about 26 kt 
matches the
mass of Super-Kamiokande, while its superior L/E resolution allows the 
reconstruction of the oscillation pattern.                           
The detector consists of two modules with dimensions of 
$14.5 \times 15 \times 13$ m$^3$ each (fig. \ref{fig:modetect}). 
Each module is made of 125 horizontal iron
layers (8 cm thick) interleaved with active planes of Glass Spark Counters
(Glass RPC's), and surrounded by external scintillation counters to help
reducing the background from cosmic ray muons.         
A magnetic field of 1.3 T, fully contained within the iron plates, is 
applied. Optionally, and end cap of vertical planes could improve the 
performance for auxiliary measurements  in the CERN to Gran Sasso beam.
If approved promptly, the experiment could start data taking towards 
the end of 2004.

Finally, a similar detector concept is also discussed in the context of a
50 kt detector for a future neutrino factory \cite{nufacdet}. Such a detector
would have a performance on atmospheric neutrinos similar to MONOLITH, but
with a larger mass, and on a longer timescale.  

\subsection{Liquid Argon TPC's: ICARUS}

Liquid Argon Time Projection Chambers (TPC's) are a further technique well 
suited to the detection of atmospheric neutrinos. Their advantage with 
respect to massive iron detectors is their sensitivity to both electron 
and muon neutrinos down to the lowest possible energies for atmospheric
neutrinos, and their sensitivity to $\tau$ appearance. Their disadvantage 
is the much higher price per kton.
The proposed ICANOE detector \cite{ICANOE}, combining a 7.6 kt (5.6 kt fiducial)      
liquid Argon TPC (ICARUS, split into 4 modules) with a 3.2 kt        
iron calorimeter (NOE), is described 
in a separate contribution to this conference \cite{LNGS-nu2000}.   
This detector is designed both for the detection of $\nu_\tau$ and $\nu_e$
appearance in the CERN to Gran Sasso neutrino beam \cite{CNGS} and for the
detection of atmospheric neutrinos of all flavours.
For some atmospheric neutrino measurements its better resolution with 
respect to Super-Kamiokande 
compensates the reduced statistics due to the lower mass.
   
A 600 t ICARUS module has already been approved and is scheduled to start
data taking in 2001. If approved promptly, the first (out of 4) ICANOE
modules could be operational in 2003.
A future 30 kt version (Super-ICARUS) has also been considered at the Letter of
Intent level \cite{Super-ICARUS}.

\section{Detection of the oscillation pattern}

\begin{figure}[ht]
\vspace{-1cm}
\epsfig{file=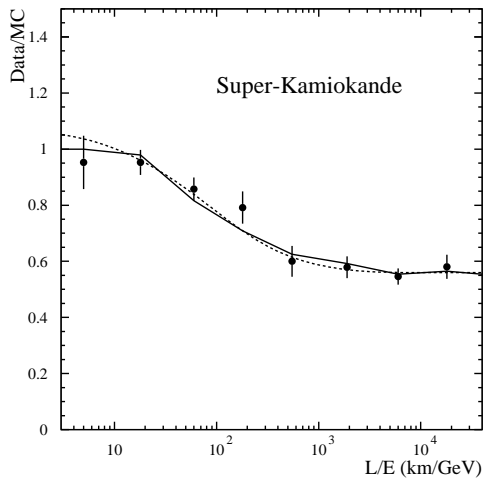,width=7cm}
\vspace{-1cm}
\caption {$L/E$ distribution from
Super-Kamiokande \cite{SK-98,SK-nu2000}  compared to the best fit oscillation
hypothesis (continous line), and to a
parametrization corresponding to the neutrino decay model of ref.
\cite{decay} (dashed line).}
\label{fig:SKLE}
\vspace{-.5cm}
\end{figure}

If detected with sufficiently high resolution, the observation of
neutrino oscillations should yield a   sinusoidal oscillation pattern 
(eq. (\ref{prob})).
However, none of the experiments which have yielded indications for
neutrino oscillations have so far succeeded to          
measure such a
pattern.  Figure \ref{fig:SKLE}   shows the
$L/E$ distribution obtained  by Super-Kamiokande \cite{SK-98,SK-nu2000}
compared
to the expectation for neutrino oscillations and to a functional form
suggested by a recent neutrino decay model \cite{decay}.  Once the
detector resolution is taken into account, the two hypotheses are
essentially indistinguishable \cite{decay}.  Even though the current
evidence is very suggestive of neutrino oscillations, a more precise
measurement of the oscillation pattern is the only way to actually
prove the oscillation hypothesis for atmospheric neutrinos. 
In particular, it should be proven 
that muon neutrinos do not only
disappear, but actually reappear at some larger L/E.

\begin{figure}[ht]
\hspace{2.8cm} 
MONOLITH 4 years
         
\vspace{-1.5cm}  
\epsfig{file=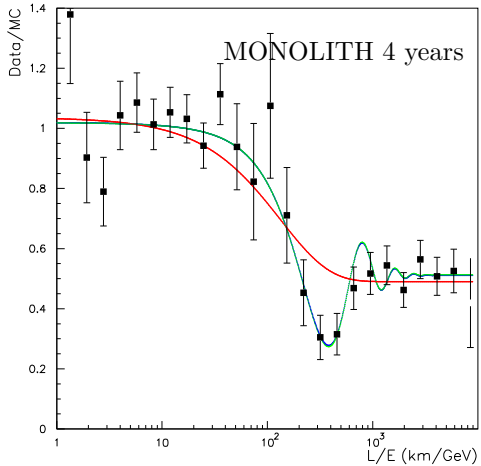,width=7cm}
\vspace{-1cm}
\caption {
$L/E$ distribution to be expected from MONOLITH \cite{MONOLITH} for
$\Delta m^2 = 3 \times 10^{-3}$ eV$^2$ compared to the best fit
oscillation hypothesis (oscillating line) and to the corresponding
best fit of the neutrino decay model of ref. \cite{decay} (smooth threshold
effect).}
\vspace{-.5cm}
\label{fig:oscdip}
\end{figure}

The proposed MONOLITH experiment \cite{MONOLITH} is explicitly designed to achieve this
goal. Having a similar mass as Super-Kamiokande, significantly larger
acceptance at high neutrino energies and better $L/E$ resolution, the
experiment is optimized to observe the full first oscillation swing,
including $\nu_\mu$ ``reappearance''.
Therefore, the oscillation hypothesis can be clearly distinguished
from other hypothesis which yield a pure threshold
behaviour (figure \ref{fig:oscdip}).

\begin{figure}[p]
\vspace{-.5cm}
\epsfig{file=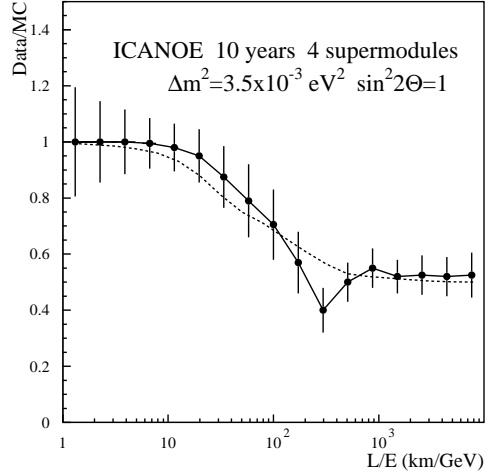,width=7cm}
\vspace{-1cm}
\caption {
$L/E$ distribution to be expected from ICANOE \cite{ICANOE} for
$\Delta m^2 = 3.5 \times 10^{-3}$ eV$^2$ compared to the best fit
oscillation hypothesis (continous line) and to the corresponding
best fit of the neutrino decay model of ref. \cite{decay} (dashed line).}
\vspace{-.5cm}
\label{fig:icanoe}
\end{figure}

A similar measurement is also being pursued by the ICANOE experiment  
(figure \ref{fig:icanoe}) and, on a somewhat longer timescale,
by the AQUA-RICH project \cite{AQUA-RICH}.

\begin{figure}[p]
\vspace{-.5cm}
\epsfig{file=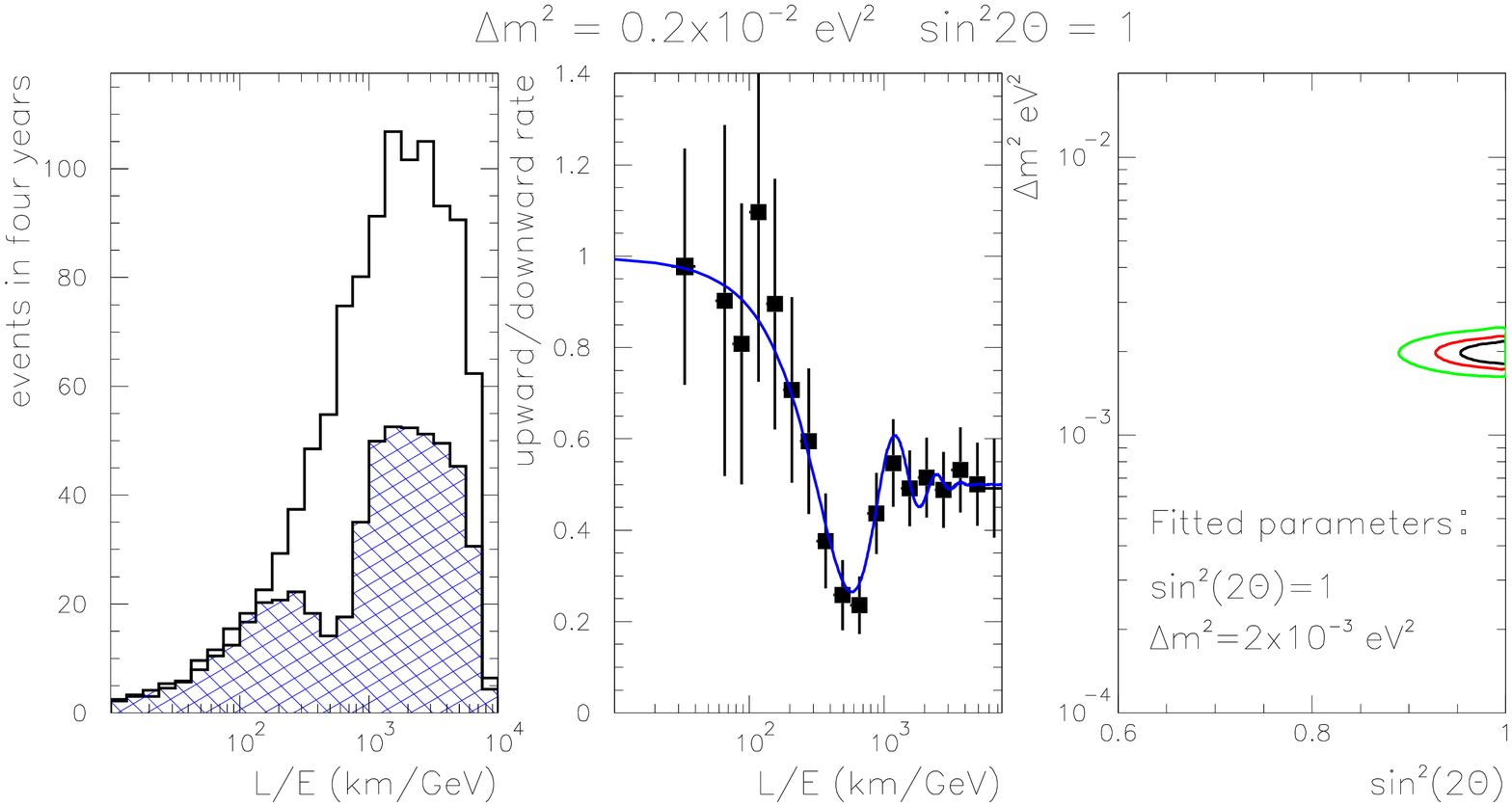,width=8cm}  
\vspace{-1cm}
\caption {
$L/E$ distribution to be expected from MONOLITH \cite{MONOLITH} for
$\Delta m^2 = 2    \times 10^{-3}$ eV$^2$.
The figures show from left to right: %
  the $L/E$ spectrum of upward muon neutrino events (hatched area) and
  the $L/E$ ``mirrored'' spectrum of downward muon neutrino events
  (open area); their ratio with the best fit superimposed; 
   and the result of the
  fit with the corresponding allowed regions for oscillation
  parameters at 68\%, 90\% and 99\% C.L.. } 
\vspace{-.5cm}
\label{fig:dm002}         
\end{figure}

Furthermore, the sensitivity to detect the oscillation pattern does 
not strongly depend on 
the oscillation parameters.
This is in contrast to long baseline experiments like MINOS,
which can do a similar measurement at the highest allowed $\Delta m^2$
if the low energy beam is used
\cite{MINOSmin}, but for which the observation of a reappearance signal in
the lower $\Delta m^2$ range is not obvious.

Fig. \ref{fig:dm002} shows the expected pattern in MONOLITH for 
$\Delta m^2 = 2 \times 10^{-3}$ eV$^2$. This figure also illustrates
the result of normalizing the sample of upgoing (oscillated) neutrinos
to the sample of downward going (unoscillated) ones.
Taking the ratio eliminates most of the systematic error due to the 
atmospheric neutrino flux and the detector response 
\cite{MONOLITH,goodLE}.

\begin{figure}[htb]
    \begin{center}
    \vspace{-.8cm}
    \epsfig{file=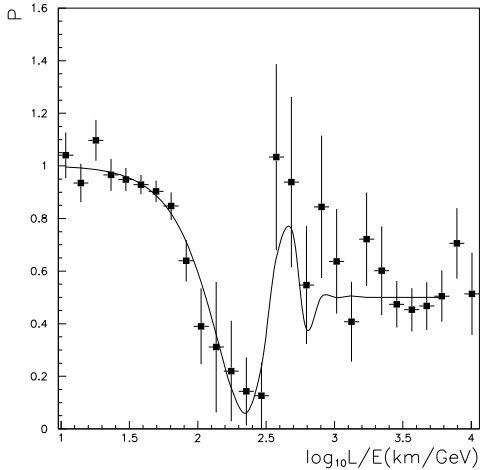,width=7cm}         
    \vspace{-.8cm}
    \caption {Example for the expected $L/E$ distribution ($\nu_\mu$
     survival probability) in MONOLITH
     with 2 years of data taking of atmospheric neutrinos and 1 year with
     the CERN-Gran Sasso neutrino beam for
     $\Delta m^2 = 5\times 10^{-3}$ eV$^2$ (points) \cite{MONOLITH}.
     Also shown is a fit of the expected oscillation hypothesis
     (continous line).
     Beam neutrinos dominate the L/E region
     below 10$^2$ km/GeV. For L/E $> 10^2$ km/GeV only
     atmospheric neutrinos contribute. Only statistical errors are shown.}
     \vspace{-.5cm}
   \label{fig:monbeam}    
    \end{center}
\end{figure}

Fig. \ref{fig:monbeam} gives an example how the atmospheric neutrino
measurement could be complemented by a disappearance measurement in
a high energy          neutrino beam. This possibility is particularly
attractive for high $\Delta m^2$ (here: $5 \times 10^{-3}$ eV$^2$).

\begin{figure}[p] 
\begin{center}
\vspace{-1cm}
\epsfig{file=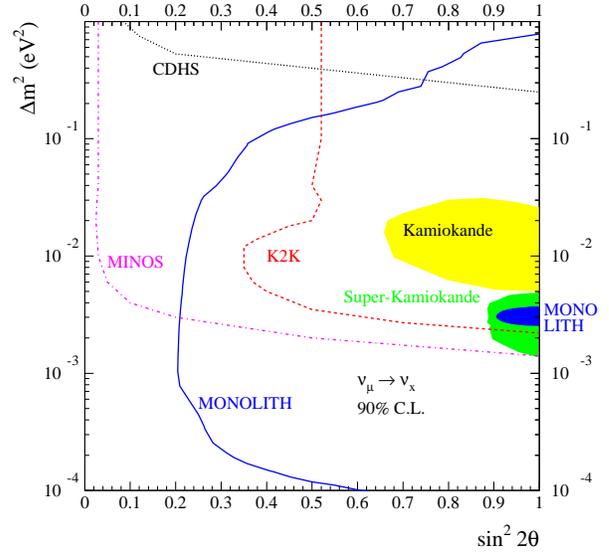,width=7.8cm} 
\end{center}
\vspace{-1.2cm}
\caption {Expected sensitivity contours for $\nu_{\mu}$ disappearance 
for K2K 
\cite{K2K,K2Knew} (5 years), MINOS \cite{MINOS} (3 years reference beam)
and MONOLITH \cite{MONOLITH} (4 years).
Also shown are the exclusion contour from CDHS \cite{others},
the recent allowed areas
from Super-Kamiokande \cite{SK-nu2000}   and Kamiokande
\cite{Kamiokande} and the expected allowed area of MONOLITH (dark shaded, 
four years) for the
current Super-Kamiokande central value.}
\label{fig:dmsincont}
\vspace{-0.5cm}
\end{figure}

\begin{figure}[p]
\vspace{-1cm}
\epsfig{file=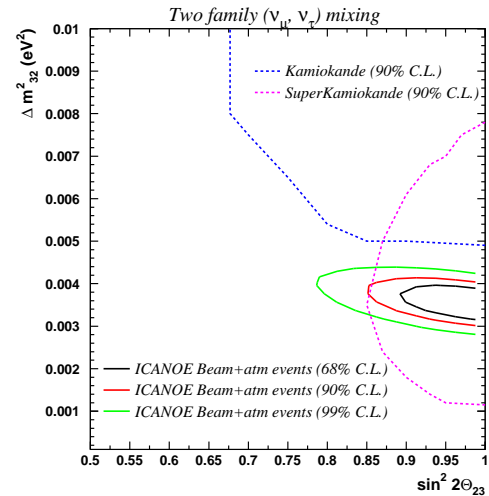,width=7cm}
\vspace{-1cm}
\caption {Overall combined fit (beam + atmospheric, appearance + disappearance)
expected by ICANOE \cite{ICANOE}, compared to Kamiokande \cite{Kamiokande} 
and Super-Kamiokande \cite{SK-98}.}
\vspace{-.5cm}
\label{fig:ICAbeamatm}
\end{figure}

Finally, the pattern measurement   can be used to significantly
improve the measurement of the oscillation parameters 
(fig. \ref{fig:dmsincont}).
Again, the result could potentially be improved by combining beam 
measurements, including the potential $\tau$ appearance rate, with 
atmospheric measurements (fig. \ref{fig:ICAbeamatm}).
In such a  combination, 
$\sin^2 2\theta$ is mainly constrained by the atmospheric
neutrino measurement, which allows a precise $\Delta m^2$ determination
from the beam information if $\Delta m^2$ is not too low.

\section{Detection of $\nu_\tau$ appearance}

For maximal $\nu_\mu - \nu_\tau$ oscillations in the $\Delta m^2$
range $2-5 \times 10^{-3}$ eV$^2$, the expected $\tau$ appearance
rate in atmospheric neutrinos is about 1 per kty \cite{ICANOE}.

Such events could be directly observed in ICANOE, but the low rate
limits the statistical significance to about 2.5 $\sigma$ in 4 years
\cite{ICANOE}.
In MONOLITH the rate is larger due to the larger mass, but the $\tau$
candidates can only be detected indirectly through an apparent 
enhancement of the NC up/down ratio at high energies (hadronic decays
of upward going $\tau$'s). Again, the significance is limited to 
about 2-3 $\sigma$ in 4 years.
The best results on $\nu_\tau$ appearance are therefore expected to be
obtained from long baseline beams. However, these atmospheric
neutrino measurements could yield useful complementary information
in the case of complicated hybrid oscillation models.

To obtain a significant sample of atmospheric $\tau$ neutrinos, bigger 
detectors like Super-ICARUS \cite{Super-ICARUS} or UNO \cite{UNO} 
might be needed. 

\section{Detection of matter effects}

Matter effects can play an important role if there are significant
contributions of $\nu_e$ or $\nu_{\rm sterile}$ to atmospheric 
neutrino oscillations. Already now, Super-Kamiokande uses the 
nonobservation of large matter effects to exclude the pure 
$\nu_\mu - \nu_{\rm sterile}$ oscillation hypothesis at   99\% c.l. 
\cite{SK-nu2000}.
For a contribution of nonmaximal $\nu_\mu - \nu_{\rm sterile}$ 
oscillations, matter effects would also manifest themselves in 
differences in the oscillation patterns for neutrinos and 
antineutrinos. Such differences could be measured with MONOLITH
\cite{MONOLITH}, and could yield important constraints on 
hybrid oscillation scenarios \cite{Lisi,yasudahybrid}.

\begin{figure}[p]
\vspace{-.5cm}
\epsfig{file=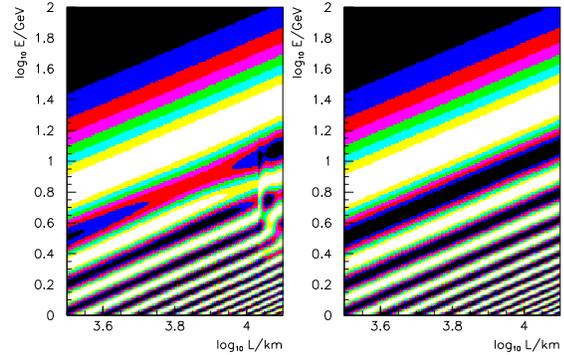,width=8.2cm}
\vspace{-1.5cm}
\caption {Muon neutrino survival probability (black=1, white=0) for three 
flavour oscillations assuming
the Super-Kamiokande ``best fit'' parameters \cite{SK-nu2000} of 
$\Delta m^2 = 2.5 \times 10^{-3}$ eV$^2$, $\sin^2\theta_{23} = 0.55$
($\sin^2 2\theta_{23} = 0.99)$ and $\sin^2 \theta_{13} = .02$
($\sin^2 2\theta_{13} = 0.08)$. {\it Left:} neutrinos. {\it Right:}
antineutrinos. Formulas are taken from an analytic two-density earth model
including core effects \cite{Chizov}. 
For negative $\Delta m^2$, the two plots would be reversed.}
\vspace{-.5cm}
\label{fig:xxx}
\end{figure}

\begin{figure}[p]
\epsfig{file=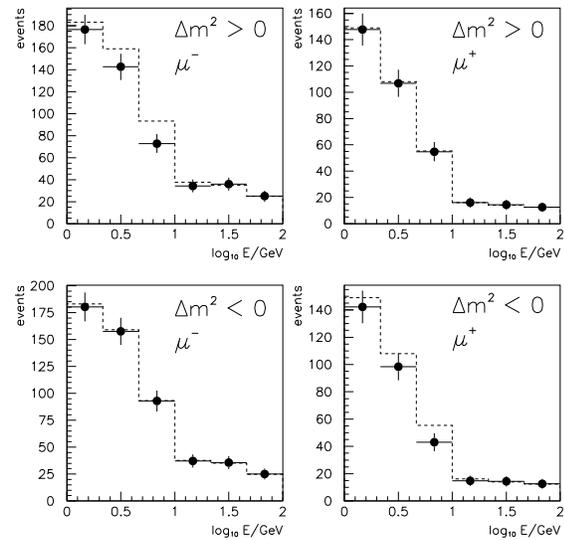,width=8.2cm}
\vspace{-1.5cm}
\caption {Reconstructed neutrino energy distributions (points) for the 
oscillation parameters of fig. \ref{fig:xxx}, for a 50 kt magnetized iron 
detector (5 years of atm. $\nu_\mu + \nu_e$). Only the events crossing 
the inner mantle 
($3200 < L < 10600$ km) are used here. The dashed line 
shows the same distribution for $\sin^2 \theta_{13}=0$ 
($\nu_\mu -\nu_\tau$ mixing only).}
\vspace{-.5cm}
\label{fig:yyy}
\end{figure}

Interestingly, such effects could be detectable even in standard three   
flavour oscillation scenarios (Figs. \ref{fig:xxx} and \ref{fig:yyy}).
A small $\nu_\mu-\nu_e$ contribution, close to the CHOOZ limit
\cite{CHOOZ}, could be strongly enhanced in 
particular regions of phase space through 
matter resonances, such that it could be measured in $\nu_\mu$
disappearance. Depending on the sign of $\Delta m^2$, such 
enhancements occur either for neutrinos or for antineutrinos only.
By comparing the neutrino and antineutrino distributions in a 
magnetized iron detector, the sign
of $\Delta m^2$, and therefore the neutrino mass hierarchy, 
could be determined if a signal would be observed.
Even a low statistics measurement could be used for this purpose if
the $\nu_\mu - \nu_e$   oscillation parameters were already measured by
the forthcoming long baseline neutrino experiments  for which matter 
effects are essentially negligible.  

Finally, the  ICANOE experiment, which is able to detect atmospheric
$\nu_e$'s down to the lowest possible energies (but not their charge),
has a window of opportunity \cite{ICANOE} to detect a subleading 
$\nu_\mu -\nu_e$
contribution from the ``solar'' $\Delta m^2$ if the large angle 
MSW solution of the solar neutrino problem \cite{solar} would turn
out to be true. Such a contribution would also be strongly affected
by matter effects.

\section{Conclusion and acknowledgements}

In conclusion, several experiments are in preparation which could
improve the present atmospheric neutrino measurements.
Such measurements should prove (or disprove) the oscillation hypothesis
for atmospheric neutrinos by measuring the oscillation pattern, 
contribute to disentangle potential 
complicated hybrid oscillation solutions from the simple 
$\nu_\mu -\nu_\tau$ oscillation scenario, and potentially
determine the neutrino mass hierarchy through the observation of
matter effects even in standard three neutrino scenarios.             

The author would like to thank the organizers of the Neutrino 2000
conference for their hospitality, and is endebted to C.K. Jung, A. Rubbia, 
and K. Zuber for providing important material for this talk.

\end{document}